\documentclass[twocolumn,prl,showpacs,amsmath,amssymb]{revtex4}
\usepackage{graphicx,color}

\begin{document}
\title{Stringent bounds on the brane width from
stellar interferometry \\ and distant gamma ray bursts: Back to the
hierarchy problem?  }

\author{Michael~Maziashvili}
\email{maziashvili@hepi.edu.ge}\affiliation{Department of Physics,
Tbilisi
State University, 3 Chavchavadze Ave., Tbilisi 0128, Georgia \\
Institute of High Energy Physics and Informatization, 9 University
Str., Tbilisi 0186, Georgia }

\begin{abstract}
A simple idea restricting the brane width due to astronomical observations is
proposed. Not to contradict the observational data the brane width should be of
about Planck size giving therefore strict criterion in selecting the realistic
braneworld models.

\end{abstract}




\maketitle

If the background metric undergoes quantum fluctuations, the fluctuations will show up in
measuring the space-time intervals. Therefore, if it is principally impossible
to measure the space-time intervals precisely, this intrinsic limitation is
naturally interpreted as a result of background metric fluctuations. Hence,
the importance of carrying out the space-time measurement to evaluate the
fluctuations of the background metric cannot be over-emphasized. To measure
the distance between two points we need the clock and the mirror situated at
those points respectively. The measurement is performed by sending the light
signal from clock to the mirror where it is reflected and returns to the
clock. However, quantum fluctuations in the positions of the clock and mirror
introduce an inaccuracy in distance measurement \cite{SW}. Uncertainties
contributed by the clock and the mirror to the measurement can be accurately
evaluated \cite{Ma}. In the case of brane there is an additional source of
error in length measurement on the brane introduced by the brane width.
Usually, the standard model fields are confined to the brane within some
localization width, i.e., brane width \cite{RSA, Gab}. If the brane width is $\epsilon$ it means that brane localized particle probes
this length scale across the brane and therefore the observer can not measure
the distance on the brane to a better accuracy than $\epsilon$. This
result immediately indicates that brane localized observer can never know a time duration to a
better accuracy than $\epsilon$, (we are using the system of units
$\hbar=c=1$). The uncertainties in space-time measurements naturally produce
the uncertainties in energy-momentum measurements, for the particle with
momentum $p$ has the wavelength $\lambda=2\pi p^{-1}$ and due to length
uncertainty one finds $\delta p=2\pi \lambda^{-2}\delta\lambda,~\delta
E=pE^{-1}\delta p$. So that the energy-momentum uncertainties of a brane
localized particle caused by the brane width read \begin{equation}\label{enmomunc} \delta
p={p^2\epsilon\over 2\pi}~,~~~~~~\delta E={(E^2-m^2)^{3/2}\epsilon\over 2\pi
E}~.\end{equation} From Eq.(\ref{enmomunc}) one sees that the
energy scale $E$ for the particle with mass $m\ll E$ and the
localization width $\epsilon$ is defined with accuracy $\delta E\approx E^2\epsilon/2\pi $. In the case $\epsilon\sim$TeV$^{-1}$ as it is usually
assumed in braneworld models with large extra dimensions \cite{Gab}, for ultra
high energy cosmic rays, $E\sim 10^{8}$TeV, one gets $\delta E\sim
10^{16}$TeV. Thus, the very high energy cosmic rays place the bound $\delta
E\lesssim 10^{8}$TeV, which implies 
\[\epsilon \lesssim 3.894\times  10^{-24}\mbox{cm}~.\] The energy-momentum uncertainties due to brane width result
in uncertainties of phase and group velocities of the photon 
\begin{equation} \label{unclspee} \delta v_p={\delta E\over p}+{E\delta p\over
p^2}={E\epsilon\over\pi}~,~~~~~~\delta v_g={d\delta E\over dp}={E\epsilon\over\pi}~. \end{equation} But is such an effect observable? Let us follow an interesting idea proposed
in \cite{LH}. The idea is to consider the phase incoherence of light coming to
us from extragalactic sources. Since the phase
coherence of light from an astronomical source incident upon a two-element
interferometer is necessary condition to subsequently form interference
fringes, such observations offer by far the most sensitive and uncontroversial
test. The light with wavelength $\lambda$ traveling
over a distance $l$ accumulates the phase uncertainty
\begin{equation*}\delta\varphi={2\pi l\over \lambda}\delta\left({v_p\over
v_g}\right)={4lE\epsilon\over\lambda}~. \end{equation*} The interference pattern when the source is viewed through a
telescope will be destroyed if $\delta \varphi$ approaches $2\pi$. In other
words, if the light with wavelength $\lambda$ received from a celestial
optical source located at a distance $l$ away produces the normal interference
pattern, the corresponding phase uncertainty should satisfy the condition
\begin{equation}\label{intercond} \delta\varphi={8\pi l\epsilon\over
\lambda^2} < 2\pi~.  \end{equation} Consider the examples used in \cite{LH}. The Young's type of interference effects were clearly seen
at $\lambda = 2.2 \mu$m light from 
a source at $1.012$ kpc distance, viz. the star S Ser, 
using the
Infra-red Optical Telescope Array, which enabled a radius determination of
the star \cite{BTG}. From Eq.(\ref{intercond}) one gets \[\epsilon \lesssim 
0.385\times 10^{-29}\mbox{cm}~. \]

Airy rings  (circular diffraction) were clearly visible
at both the zeroth and first maxima in an observation
of the active galaxy PKS1413+135 ($l = 1.216$ Gpc) by
the Hubble Space Telescope at $ 1.6 \mu $m wavelength \cite{PSC}.
Correspondingly, using the Eq.(\ref{intercond}) one gets the following restriction 
on epsilon \[\epsilon \lesssim  0.169\times 10^{-35}\mbox{cm}~. \]
 
The above consideration tells us that the error made in measuring the length
$l$ by the light with wave length $\lambda$ approaches \[\delta l=\epsilon
{l\over \lambda}~,\] since the wavelength can not be known with a better
accuracy than $\epsilon$. The uncertainty in length measurement will lead to
apparent blurring of distant point sources observed through the telescope
\cite{RTG}. Considering the distances $l_1$ and $l_2$ as measured from a point source placed at a distance 
$l\approx l_1 \approx l_2$ from the two sides of a telescope of aperture $D$, any intrinsic  variation $\delta l$ in the wavefront along the two lines of sight will translate into
 an apparent angular shift $\delta \theta$ given by \cite{RTG}

\begin{equation}
\delta \theta \approx {\delta l\over D}~.
\end{equation} So one can state that the fluctuations in space-time measurements on the brane
due to brane width lead to
an apparent angular broadening of a light source placed at a distance $l$, 
as seen from a telescope of diameter $D$, given by

\begin{equation}
\delta \theta \approx {\epsilon l\over \lambda D} ~. 
\end{equation} From this point of view, the analyses of diffraction pattern from the Hubble Space Telescope
observations of SN 1994D \cite{Pat} and Hubble Deep Field high-$z$ images
\cite{Spin} puts the following restriction on the brane width \cite{RTG}
\[\epsilon\lesssim 3\times 10^{-36}\mbox{cm}~.\]

Another idea is to use experimental limits on the variation of light speed.
From Eq.(\ref{unclspee}) one sees that fluctuations of light speed increase
with energy. Thus for photons emitted simultaneously from a distant
source coming towards our detector, we expect an energy dependent spread in
their arrival times. To maximize the spread in arrival times, it is desirable
to look for energetic photons from distant sources. This proposal was first
made in another context in \cite{ACE}. The analyses of the most rapid TeV
flare observed thus far from active galaxy Markarian 421 on 15 May
1996 \cite{Gaid} puts the following limit on the brane width \cite{Bil, Sch}\[\epsilon \lesssim {\pi\over 4}\times
10^{-16}\mbox{GeV}^{-1}\approx 
0.973\times 10^{-29}\mbox{cm}~. \] 

\begin{figure}[t]

\includegraphics{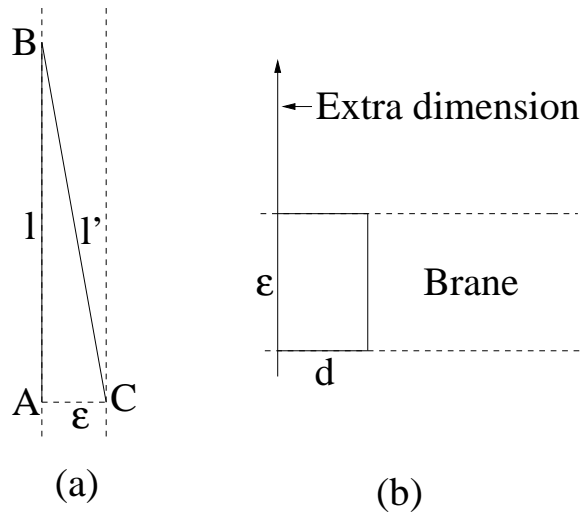}\\
\caption{ (a) The light signal traveling the path clock-mirror-clock; (b)
``Distortion'' of the clock due to brane width. }
\end{figure}

At the first glance one can think the situation is not so ``hopeless'' as it
is depicted above. Namely one can consider a schematic picture of the
measurement given in Fig.1 (a). Clock sending the light from point $A$ to the
mirror situated at distance $l$ (point $B$) can detect the signal coming back
at point $C$ (for the clock undergoes quantum
fluctuations across the brane) and therefore uncertainty in length measurement should be $l'-l\sim
\epsilon^2/l$, which is much less than $\epsilon$ for relatively large
distances $l\gg \epsilon$. Unfortunately, clock can not measure the time with
accuracy greater than $\epsilon$. Imagine the clock as a mirror box with
linear size $d$ inside which light is bouncing. In absence of extra dimensions
the resolution time of this clock will be $ d$. But in the case of
braneworld scenario, the light is localized within $\epsilon$ along the extra
dimension(s) and therefore our clock is a box $\epsilon\times d$ inside which
light is bouncing, Fig.1 (b). Hence, the accuracy of
the clock can not be greater than $\epsilon$, no matter what the length scale
$d$ is. During the time interval $\epsilon$ the clock is capable to resolve,
the light will travel a distance $\epsilon$ resulting thereby this error in
the length measurement. So, one sees that the brane width actually represents
an error in length measurement on the brane.

Strictly speaking the experimental results
considered here have to do mainly with the photon localization width, which in
general may be different from the localization widths of other species of
particles. However, if we want to address the hierarchy problem they ought not
to differ very much. But even in this case we need to explain the natural
appearance of so small brane width from higher-dimensional theory with TeV
fundamental scale. Without this attempt we are left with the hierarchy
(problem?) between the size of extra dimensions $\sim$Tev$^{-1}$ and the
brane width that should be of about Planck size.

The author is greatly indebted to David Langlois for cordial hospitality at
APC (Astroparticule et Cosmologie, CNRS, Universit\'e Paris 7) and IAP
(Institut d'Astrophysique de Paris), where this work was done. Thanks are due
to C.~Deffayet and P.~Peter for useful conversations. The work was supported by
the \emph{INTAS Fellowship for Young Scientists}, the
\emph{Georgian President Fellowship for Young Scientists} and the
grant \emph{FEL. REG. $980767$}.

\end{document}